\documentclass[aps,pra,onecolumn,showpacs,amsmath,amssymb,superscriptaddress,longbibliography]{revtex4-2}

\usepackage{bm}
\usepackage{amsthm}
\usepackage{amsfonts}
\usepackage{graphicx}
\usepackage{epsfig}
\usepackage{epstopdf}
\usepackage{subfigure}
\usepackage[colorlinks=true,citecolor=blue,urlcolor=blue]{hyperref}
\usepackage{color}
\usepackage{ulem}

\newcommand{\be}{\begin{equation}}
\newcommand{\ee}{\end{equation}}
\newcommand{\ba}{\begin{eqnarray}}
\newcommand{\ea}{\end{eqnarray}}
\newcommand{\ban}{\begin{eqnarray*}}
\newcommand{\ean}{\end{eqnarray*}}

\newcommand{\one}{\leavevmode\hbox{\small1\normalsize\kern-.33em1}}

\begin{document}

\title{Collective oscillations of a Bose-Einstein condensate induced by a vortex ring}
\author{Wen-Kai Bai}
\author{Jian-chong Xing}
\affiliation{Institute of Modern Physics, Northwest University, Xi'an 710069, China}
\author{Tao Yang}
\email{yangt@nwu.edu.cn}
\affiliation{Institute of Modern Physics, Northwest University, Xi'an 710069, China}
\affiliation{Shaanxi Key Laboratory for Theoretical Physics Frontiers, Xi¡¯an 710069, China}
\author{Wen-Li Yang}
\affiliation{Institute of Modern Physics, Northwest University, Xi'an 710069, China}
\affiliation{Shaanxi Key Laboratory for Theoretical Physics Frontiers, Xi¡¯an 710069, China}
\author{Wu-Ming Liu}
\email{wmliu@iphy.ac.cn}
\affiliation{Beijing National Laboratory for Condensed Matter Physics, Institute of Physics, Chinese Academy of Sciences, Beijing, China}
\affiliation{School of Physical Sciences, University of Chinese Academy of Sciences, Beijing 100190, China}
\affiliation{Songshan Lake Materials Laboratory, Dongguan, Guangdong 523808, China}

\date{\today}

\begin{abstract}
We study the collective oscillations of three-dimensional Bose-Einstein condensates (BECs) excited by a vortex ring. We identify independent, integrated, and stationary modes of the center-of-mass oscillation of the condensate with respect to the vortex ring movement. We show that the oscillation amplitude {of the center-of-mass of the condensate} depends strongly on the initial radius of the vortex ring, the inter-atomic interaction, and the aspect ration of the trap, while the oscillation frequency is fixed and equal to the frequency of the harmonic trap in the direction of the ring movement. However, when applying Kelvin wave perturbations on the vortex ring, the center-of-mass oscillation of the BEC is changed nontrivially with respect to the perturbation modes, the long-scale perturbation strength as well as the wave number of the perturbations. The parity of the wave number of the Kelvin perturbations plays important role on the mode of the center-of-mass oscillation of the condensate.
\end{abstract}

\maketitle
\section{\label{sec:level1}INTRODUCTION}


Vortex rings, which are fascinating phenomena exist widely in nature, from normal fluid to quantum systems, such as smoke rings, quantized vortex rings in superfluid helium \cite{PRev.136.a1194}, and optical vortices \cite{Prog.Opt.53.293}. In three-dimensional (3D) systems, vortex rings are stable nonlinear modes, which have been an active subject of experimental and theoretical researches due to their compact and persistent nature. Bose-Einstein condensates (BECs) provide an ideal platform for investigating this exotic phenomenon, due to the very well-controlled nature of condensates. Anderson {\it{et al.}} first imaged stable vortex rings experimentally which were formed by the decay of a dark soliton in a trapped two-component BEC \cite{PRL.86.2926}. The evidence that the solitons can evolve periodically between vortex rings and solitons in a elongated condensate is given in Ref. \cite{NP.5.193}. Some other methods are proposed to generate a vortex ring in a BEC, such as by means of electromagnetically induced atomic transitions \cite{PRL.86.3934}, nonlinear quantum piston \cite{PRA.87.053624}, and optical tweezers \cite{PRA.88.043637}.
The stability and dynamics of vortex rings are extensively studied in Refs. \cite{PRA.60.4882, PRA.70.033605,PRA.74.041603,PRA.74.041603,PRA.88.053626,PRA.92.063611,PRA.98.033609,NP.5.193,PRL.86.3934,PRL.94.040403}.

Collective modes play an important role in elucidating the physical properties of trapped atomic BECs and quantum degenerate Bose and Fermi gases \cite{PRL.108.035302,PRL.109.115301,PRL.91.250402,PRE.93.022214}. Measuring the collective modes can help us to understand the physics of a quantum many-body system. Dipole oscillation, which represents a center-of-mass oscillation of all atoms, is the first and simplest collective excitation mode for a harmonically trapped BEC \cite{PRA.58.2385,JLTP.115.61,JPB.47.035302}. If the Galilean invariance is valid in the system, the frequency of this mode is exactly the same as the trap frequency in the direction of the movement of the condensate, independent of the oscillation amplitude and interatomic interactions, which can be understood by the use of Kohn theorem \cite{PRev.123.1242,RMP.71.463,PRL.77.2360}. In the traditional schemes these oscillations are obtained by shaking the condensate through the modulation of the trapping magnetic fields \cite{PRL.77.988,PRL.109.115301}. However, the collective modes excited by the creation and dynamics of vortex rings has never been studied in detail. Moreover, measuring the collective modes can help us to understand the dynamics of rings better.

In a trapped BEC, a vortex ring will move in response to the effect of the non-uniform trap potential and the external rotation, as well as self-induced effects caused by its own local curvature \cite{Alexander}. Kelvin waves provide a source of perturbations for the motion of the vortex ring. The translational self-induced velocity of the vortex ring can be reduced if the ring is perturbed by helical Kelvin waves of given amplitude and azimuthal wave number \cite{PRE.74.046303,PRA.83.045601}. It was also suggested that a slightly oblate trap is needed to prevent the vortex ring from becoming unstable due to bending waves \cite{PRA.74.041603}. However, the effects of the Kelvin waves on the dipole oscillation of the condensate is still not clear.

This paper is organized as follows. In section \ref{ring} we describe the model used to construct a vortex ring in a 3D condensate system, and the model of ring dynamics. The numerical parameters and procedure for simulating the dynamics of the given system is also introduced in this section. In section \ref{dipole}, we identify the factors that influence the dipole oscillation of the center-of-mass of the condensate with a vortex ring, including the geometry of the condensate, the Thomas-Fermi (TF) radius of the condensate and the radius of the vortex ring. In section \ref{kelvin} we introduce Kelvin wave perturbations to explore its impact on the dipole oscillation of the condensate induced by the imprinted vortex ring. Section \ref{conclusion} contains our conclusions.

\section{ vortex ring }\label{ring}

At sufficiently low temperatures, the macroscopic behavior of a trapped BEC with $N$ atoms is well characterized by the Gross-Pitaeskill equation (GPE) \cite{RevModPhys.73.307},
\begin{equation}
i\hbar\frac{\partial \psi}{\partial t}=\left(-\frac{\hbar^2}{2m}\nabla^2+V+gN\left|\psi\right|^2\right)\psi\text{,}
\end{equation}
where $\psi$ is the order parameter of the condensate, and the coupling constant $g=4\pi\hbar^2a_s/m$ is related to the s-wave scattering length $a_s$ and the mass of the atoms $m$. We consider a harmonic trap 
$V=m(\omega_\perp^2x^2+\omega_\perp^2y^2+\omega_z^2z^2)/2$ with $\omega_\perp$ and $\omega_z$ being the trap frequencies in the radial and axial directions, respectively.

\begin{figure}[t]
\begin{center}
\includegraphics[angle=0,width=0.27\textwidth]{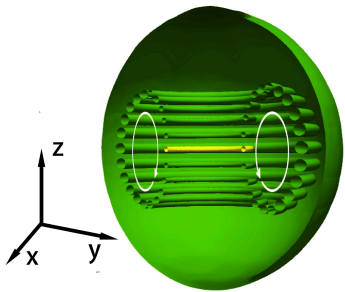}
\caption{(Color online) The time evolution of the isosurfaces of the density field for a vortex ring. The initial radius of the vortex ring is $R_{v}=2a_0$, and the ring is located at $z=0$ plane at $t=0$ as shown by the yellow ring. The direction of the movement is indicated by the white circles. The parameters chosen are $\lambda_z=1$ and $N=1.5\times10^5$.}
\label{fig1}
\end{center}
\end{figure}

A vortex ring can be realized by means of a loop vortex line similar to the method used to create vortex knots in Ref. \cite{PRE.85.036306}. Suppose a vortex ring is initially placed on the $xoy$ plane and rotationally symmetric about the $z$ axis, it can be treated as an assemble of 2D vortex dipole on the $roz$ plane with $r=\left|\mathbf{r}\right|=\sqrt{x^2+y^2}$. To imprint a vortex core located at $(\mathbf{r}_j,0)$ in the $roz$ plane, the phase space configuration is
\begin{align}
\theta(r-r_j,z)=s_j\text{atan2}(z, r-r_j)
\label{phase}
\end{align}
where atan2(...) is the extension of the arctangent function whose principal value is in the range $(-\pi, \pi]$, and $s_j$ is the topological charge of the vortex. Due to the symmetry of the system, the wave function of the condensate with a vortex ring can be obtained by
\begin{equation}
\psi_{3D}=\sqrt{\rho}\text{exp}\left\{i\theta[r(x,y)-R_v, z]-i\theta[r(x,y)+R_v,z]\right\}.
\end{equation}
{By evolving the GPE in imaginary time while enforcing a $2\pi$ winding of the phase around the location of vortex cores, we prepare the initial state of a 3D condensate containing a vortex ring with radius $R_v$ as shown by the yellow ring in Fig.\,\ref{fig1} for the real time propagation}.{We note that there are other methods that one can use to prepare the initial state of the system, such as by real time evolution to imprint a topological defect into the condensate \cite{PRA.87.023603, PRA.88.053626}, and by a trail wave function which is the combination of the ground state wave function and the phase and density factors \cite{PRA.88.043637, PRA.92.053603}. However, the preparation method will not change the intrinsic physics of the system. }

The ring dynamics are influenced generally by both the non-uniform trap potential (density distribution of the condensate) and the self-induced effects due to the local curvature of the ring. For a vortex ring in a homogeneous Bose fluid, one can define the self-induced velocity $v_{in}=\left(\hbar/2mR_v\right)\left[\ln \left(8R_v/\xi\right)-0.615\right]$, which propagates along the direction normal to the plane of the ring \cite{Roberts}. If the ring is created in a trapped 3D BEC and the radius of ring is less than the TF radius of the condensate, the Magnus force caused by the harmonic trap leads to precession of the ring \cite{PRA.61.013604,PRA.88.053626}. {The nonlinear dynamics of the vortex ring is explored by utilizing two numerical methods, the Crank-Nicolson scheme and the fourth-order Ronger-Kutta method, which give quantitative agreement with each other.}

In our simulations, the trap frequency in the radial direction is always $\omega_\perp=2\pi\times 75 ~\text{Hz}$, the trap aspect ratio is $\lambda_z=\omega_z/\omega_\perp$, and the bulk s-wave scattering length is $a_s = 5.4 \,\text{nm}$ for a {$^{87}$\it{$Rb$}} BEC. We use $a_0=\sqrt{\hbar/m\omega_\perp}$ and $t_0=1/\omega_\perp$ as the unit of length and time, respectively. {In the calculations, $150\times 150\times 150$ grids with steps $\triangle x=\triangle y=\triangle z=\sqrt{2}a_0/10$ are employed in a uniformly discretized physical space. A small time step, $\Delta t=0.001t_0$, is chosen to ensure accuracy of the results.}

\begin{figure}[tbp]
\includegraphics[angle=0,width=0.5\textwidth]{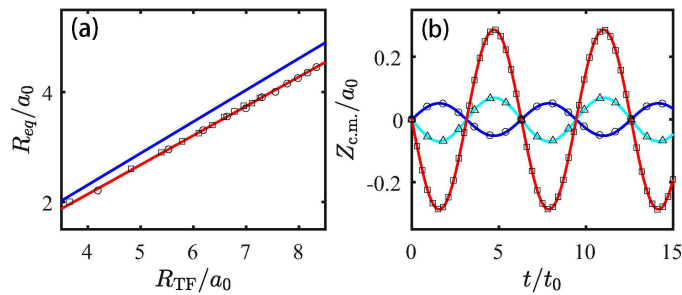}
\caption{(Color online) (a) The curves of the unstable equilibrium ring radius $R_{eq}$ as a function of the TF radius $R_{TF}$ in the radial direction from analytical (blue line) and numerical (red line) calculations. The open squares and circles are data for the trap aspect ratio $\lambda_z=1$ and 2, respectively. (b) The time evolution of the center-of-mass oscillation of the condensate in the $z$-direction with $\lambda_z=1$. The solid lines with open symbols show the $Z_{\rm {c.m.}}$ trajectories for the initial radius of the vortex ring being $R_{v}/a_0=1$ (blue line with circles), 2 (cyan line with triangles) and 3 (red line with squares), respectively. {The total number of particles are kept to be $N=4.0\times 10^4$.}}
\label{fig2}
\end{figure}

\begin{figure}[b]
\begin{center}
\includegraphics[angle=0,height=0.45\textwidth]{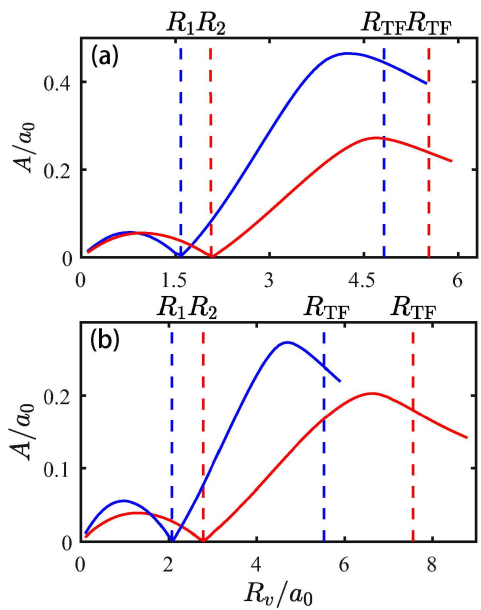}
\caption{(Color online) The amplitude $A$ of the center-of-mass oscillation of the condensate versus the initial radius $R_v$ of the vortex rings, with the blue solid lines and the red solid lines correspond to the trap aspect ratios $\lambda_z=1$ and $\lambda_z=2$ with fixed $N$ (a) and the total particle number $N=4.0\times10^4$ and $N=1.9\times10^5$ with $\lambda=2$ (b), respectively. With the special initial radius of the vortex ring $R_1$ ($R_2$), there is no center-of-mass oscillation {of the condensate} ($A=0$).}
\label{fig3}
\end{center}
\end{figure}

Figure\,\ref{fig1} shows a representative dynamics of a vortex ring with {initial} radius $R_{v}=2a_0$ in a trapped spherical BEC. 
The time evolution of the vortex ring whose initial position is denoted by the yellow ring displays a periodic oscillatory motion along the arrow direction. The oscillation is shown by taking a series of snapshots from $t=0$ to $t=20t_0$ with time interval being $t_0$. Clockwise or anticlockwise motion of the ring is identified by the topological charge distribution. The ring motion can be interpreted as a trajectory around an energy maximum at $r=R_{eq},z=0$, which can be obtained by approximately taking a ring energy that equals to a sum of a circle of energy of single 2D vortex \cite{PRA.61.013604}. It is analogous to the dynamics of a two-dimensional vortex dipole in any $roz$ plane \cite{PRA.77.053610,PRA.84.011605,SR.6.29066}.

There is a unique radius $R_{eq}$ at $z=0$ plane, corresponding to the position where the vortex ring is in the unstable equilibrium state with respect to the center-of-mass of the condensate. In the TF limit, the analytical value is $R_{eq}\approx R_{TF}/\sqrt{3}\approx0.577R_{TF}$ \cite{PRA.61.013604}, with $R_{TF}$ being the TF radius in the radial direction. For a trapped BEC, $R_{eq}$ can also be obtained by means of the asymptotic prediction in the TF limit \cite{PRA.62.063617} and the particle picture method \cite{PRA.95.043638}. Figure \ref{fig2}(a) depicts the comparison between the numerical solutions of GPE (red line) and the analytical solution (blue line) of $R_{eq}$. Our result is $R_{eq}=0.535R_{TF}$ as shown by the red solid line in Fig.\,\ref{fig2}(a), which is very close to the numerical value of $R_{eq}=0.54R_{TF}$ in Ref.\,\cite{PRA.61.013604}. It indicates the fact that the analytical solution tends to over-estimate the unstable equilibrium radius of vortex rings. The open squares and circles in Fig.\,\ref{fig2}(a) are data for $\lambda_z=1$ and 2, respectively. It is clear that the change of the aspect ratio of the trap has little effects on $R_{eq}/R_{TF}$. 

\begin{figure}[t]
\begin{center}
\includegraphics[angle=0,height=0.45\textwidth]{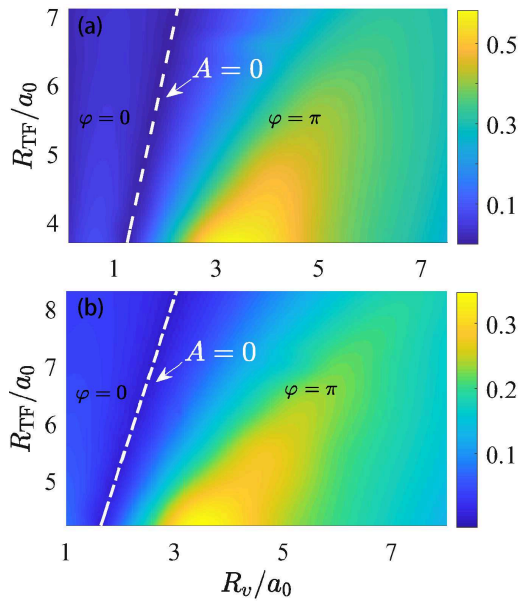}
\caption{(Color online) The amplitude $A$ of the center-of-mass oscillation as a function of $R_{TF}$ and $R_v$ with $\lambda_z=1$ (a) and $\lambda_z=2$ (b), respectively. The color shows the magnitude of $A$. The white dashed line indicates the parameter range which makes $A$ to be zero.}
\label{fig4}
\end{center}
\end{figure}

\section{Dipole oscillation of the center-of-mass of condensates}\label{dipole}

When imprinting a $2\pi$ phase to construct a vortex ring in a trapped BEC, the collective modes of the low-energy excitations are triggered by the movement of the vortex ring along the $z$-direction. The radial oscillation of the ring does not lead to any collective excitation modes for center-of-mass of the condensate in the $\mathbf{r}$ direction even though the vortex ring expands and shrinks back and forth in the radial direction. This is due to the symmetry of the system in the radial direction. The center-of-mass oscillation is then defined as
\begin{align}
Z_{\rm{c.m.}}(t)=\frac{1}{N}\int z \rho(\mathbf{r},z,t)\,d\mathbf{r}dz.
\end{align}
Figure\,\ref{fig2}(b) shows the oscillation of $Z_{\rm {c.m.}}$ as a function of time. We can see clearly that the amplitude and phase vary with $R_v$, while the frequency is not changing. All the oscillation curves can be fitted by the sinusoidal function
\begin{align}
Z_{c.m.}(t)=A\text{sin}(\omega_c t+\varphi){,}
\label{zcom}
\end{align}
with $A$ and $\omega_c$ being the amplitude and the frequency of the center-of-mass oscillation, respectively.

Through the calculations,we find that {if there are center-of-mass oscillations,} the frequencies of these collective modes always equal to the trap frequency in the $z$-direction, i.e., $\omega_c=\omega_z$, no matter what are the values of $R_v$ and $\lambda_z$, which indicates that the oscillation of the condensate in the axial direction induced by the vortex ring is the first excited state of the condensate, i.e., the center-of-mass dipole oscillation \cite{PRA.58.2385,JLTP.115.61,JPB.47.035302}.

\begin{figure}[t]
\includegraphics[angle=0,width=0.5\textwidth]{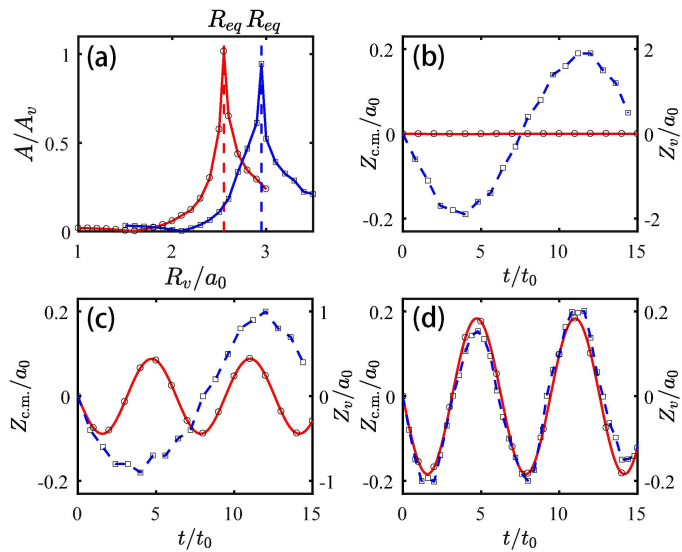}
\caption{(Color online) {(a) The ratios of the amplitude of the center-of-mass oscillation of the condensate to that of the vortex ring ($A/A_{v}$) versus the initial radius of the vortex ring with two trap configurations $\lambda_z=1$ (red solid line with circles) and $\lambda_z=2$ (blue solid line with squares). (b)-(d) Position of the center-of-mass $Z_{\rm {c.m.}}$ (red solid lines with circles) and the position of vortex ring $Z_{v}$ (blue dashed lines with squares) in the $z$-direction. The parameters chosen are $R_{v}/a_0=1.57, 2.1, 2.55$ for (b), (c) and (d), respectively, while $\lambda_z=1$ and $N =4.0\times10^4$. The open circles in (b), (c) and (d) are numerical data.}}
\label{fig5}
\end{figure}

\begin{figure}[b]
\includegraphics[angle=0,width=0.35\textwidth]{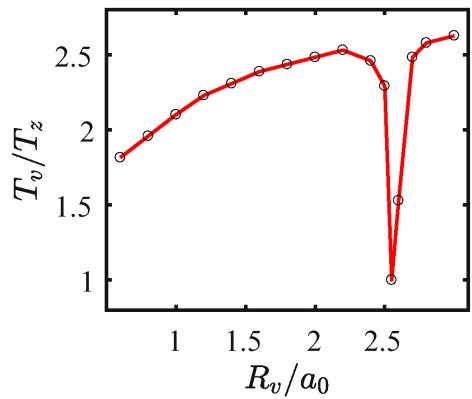}
\caption{(Color online) {Vortex ring oscillation period $T_v$ (normalized by the trap period $T_z$) as a function of the initial ring radius $R_v/a_0$. The parameters chosen are $\lambda_z=1$ and $N =4.0\times10^4$.}}
\label{fig6}
\end{figure}

The trap aspect ratio $\lambda_z$, the interatomic interaction, and the initial radius of the vortex ring $R_v$ are three critical factors that affect the oscillation amplitude of the center-of-mass oscillation of the condensate. By keeping the number of atoms a constant, the amplitude of the oscillation $A$ varies with $R_v$ for $\lambda_z=1$ (the blue line) and $\lambda_z=2$ (the red line) as shown in Fig.\,\ref{fig3}(a). For a given $\lambda_z$ but varying number of atoms, the $A-R_v$ curves are shown in Fig.\,\ref{fig3}(b). For each case, there exists a $R_v$ {($R_1$ and $R_2$ as shown in Fig.\,\ref{fig3})} at which the center-of-mass of the condensate will stay at rest even though the vortex ring is still moving inside the condensate cloud. We note that the equilibrium point for the center-of-mass of the condensate and that for the vortex ring are not in the same value. The larger $\lambda_z$ ($N$) is, the lager $R_v$ is for $A=0$ as the TF radius increases with increasing $\lambda_z$ ($N$). We expect that the stronger center-of-mass oscillation may cause stronger instability of the vortex ring. In Fig.\,\ref{fig4}, we plot the amplitude of the oscillation $A$ as a function of the TF radius $R_{\rm {TF}}$ and the initial radius of the vortex ring $R_v$ for two different trap aspect ratio $\lambda_z=1$ and 2, illustrating a very fascinating linear relation of $A=0$ in both cases as shown by the white dashed lines in the plots. From the left-hand-side to the right-hand-side region of $A=0$, there is a phase shift from $\varphi=0$ to $\varphi=\pi$ for the oscillation described by Eq.\,(\ref{zcom}).

\begin{figure}[t]
\includegraphics[angle=0,width=0.45\textwidth]{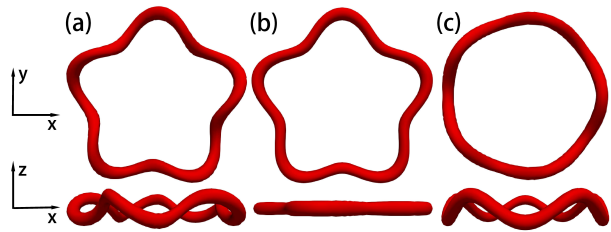}
\caption{(Color online) The top and side views of the isosurfaces of the vortex rings perturbed by helical (a), planar (b) and vertical (c) Kelvin waves with the wave number being $n=5$ the relative amplitude being $B_{xy}=B_z=0.3a_0$. The inial radius of the vortex ring is $R_v=2a_0$, and the number of atoms is $N =1.0\times10^5$. }
\label{fig7}
\end{figure}

The relation between the ratio of the oscillation amplitude of the center-of-mass of the condensate and that of the vortex ring's movement in the $z$-direction with respect to the initial radius of the vortex ring is shown in Fig.\,\ref{fig5}(a). We can see that when the initial radius of the vortex ring is far from $R_{eq}$ marked by the dashed lines, the value of $A$ is much smaller than that of the oscillation amplitude of the vortex ring, $A_{v}$, indicating that the weak influence of the center-of-mass oscillation on the vortex ring dynamics. In Figs.\,\ref{fig5}(b)-(d), we show that the time evolution of the position of the vortex ring and the center-of-mass of the condensate in the $z$-direction, $Z_v$ and $Z_{c.m.}$, {with three typical values $R_{v}/a_0=1.57, 2.1, 2.55$, which discloses three distinct dynamical regimes of the center-of-mass oscillation and the vortex ring oscillation. As stated before, for a specific value of $R_v$ ($R_1$ and $R_2$ for $\lambda=1$ and $\lambda=2$, respectively, in Fig.\,\ref{fig3}), the center-of-mass of the condensate cloud can stay stationary while the vortex ring oscillates inside the condensate as shown in Fig.\,\ref{fig5}(b). These two oscillations can also evolve with different amplitude and frequency as shown in Fig.\,\ref{fig5}(c), as if two oscillations are independent.} In the case {shown in Fig.\,\ref{fig5}(d) with $R_v=R_{eq}$}, the motion of the vortex ring {(blue dashed line)} is integrated with the center-of-mass oscillation of the condensate (red solid line) with the same frequency and nearly the same amplitude. In this case, the instability of the ring mainly depends on the center-of-mass dipole oscillation of the condensate.

{In Fig.\,\ref{fig6}, we show the the radius dependence of the vortex ring dynamics. we find that, for a given range of $R_v$, the rings with smaller $R_v$ have smaller periods, which is consistent with the similar results reported in Refs.\,\cite{PRL.112.025301,PRA.88.053626}. However, in the vicinity of the unstable equilibrium vortex ring radius $R_{eq}$, this trend is not satisfied. The normalized oscillation period of the vortex ring, $T_v/T_z$ with $T_z=2\pi/\omega_z$, drops sharply to 1 from both sides of $R_{eq}$.}

\section{Role of the Kelvin waves}\label{kelvin}

The Kelvin waves modify the vortex ring with periodic distortions. We argue if this kind of distortions affects the dipole oscillation of the condensate. 
Let us initialize the three-dimensional wave function of the condensate with a perturbed ring as
$\psi_{3D}=\psi_{2D}\left\{l\left(x,y\right)-R_0, z-\xi\right\}\times\psi_{2D}^*\left\{l\left(x,y\right)+R_0,z-\xi\right\}$, with $l\left(x,y\right)=\sqrt{x^2+y^2}-B_{xy}sin(n\beta)$, $\xi=B_{z}cos(n\beta)$ and $\beta=atan2(y, x)$, where $B_{xy}$, $B_z$ and $n$ are the amplitude in the radial and axial directions, and the wave number of the helical Kelvin wave perturbations, respectively. Figure\,\ref{fig7}(a) shows the initial state of the vortex ring with radius $R_v=2a_0$ disturbed by the helical Kelvin waves with relative amplitude $B_{xy}/R_v=B_z/R_v = 0.15$ and the wave number $n=5$. We can see clearly that helical deformations are just a combination of deformations in the radial direction and the axial direction. So we can easily create planar Kelvin waves (in the $xOy$ plane) and vertical ones (in the $z$-direction) as shown in Figs.\,\ref{fig7}(b) ($B_{xy}/R_v= 0.15$ and $B_z =0$) and \ref{fig7}(c) ($B_z/R_v =0.15$ and $B_{xy} =0$  ).

\begin{figure}[t]
\includegraphics[angle=0,width=0.45\textwidth]{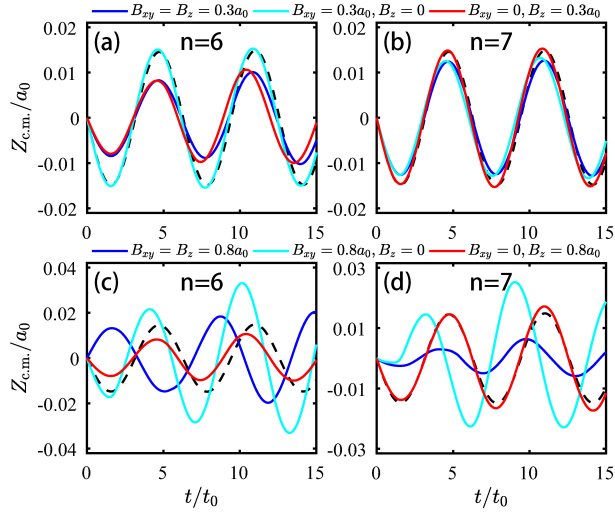}
\caption{(Color online) The trajectories of the center-of-mass oscillation of a condensate perturbed by different types of Kelvin waves. The blue, cyan and red solid lines corresponds to the cases with helical, planar and vertical Kelvin wave perturbations, respectively. The black dashed lines indicates the cases without any perturbations. The wave number of the Kelvin waves in (a) and (c) is 6, while it is 7 in (b) and (d).  }\label{fig8}
\end{figure}

For the cases when the ratio of the perturbation amplitude to the radius of the vortex ring, $B/R_v$ ($B_{xy}$ and $B_z$), is relatively small, the change of the frequency of the dipole mode is not obvious as shown in Figs.\,\ref{fig8}(a) and \ref{fig8}(b). The helical Kelvin wave perturbations depress the center-of-mass oscillation of the condensate. However, the magnitude of this depression for even $n$ is much larger than that for odd $n$. Moreover, the source of this depression also depends strongly on the parity of the wave number of the Kelvin wave perturbations. For $n$ is even, we can see clearly that the change of the amplitude of the center-of-mass oscillation of the condensate is mainly from the contribution of the vertical Kelvin waves as shown in Fig.\,\ref{fig8}(a) by the red solid line. If there are only perturbations in the $xoy-$ plane, the oscillation mode of the center-of-mass of the condensate is nearly identical to that of the case without Kelvin waves (see the cyan solid line and the black dashed line in Fig.\,\ref{fig8}(a)). However, for cases with odd $n$, the result is opposite. The planar Kelvin waves contribute to the change of the amplitude of the oscillation as shown in Fig.\,\ref{fig8}(b).

When we increase the amplitude of the Kelvin waves, both the frequency and amplitude of the center-of-mass oscillation will deviate from the values without Kelvin waves as shown in Figs.\,\ref{fig8}(c) and \ref{fig8}(d). We find that the perturbations in the $xoy$ plane enhance the center-of-mass oscillation and also modulate the frequency of the oscillation strongly, while those in the $z$-direction only had little effects on the varying of the amplitude and the frequency of the dipole oscillation. For large $B$ ($B_{xy}$ and $B_z$), the amplitude ($A$) and the frequency ($\omega_c$) of the center-of-mass oscillation are not constants anymore, but functions of time, which indicates the breakdown of the dipole excitation mode induced by the vortex ring when the Kelvin wave perturbations break the ring structure.

\begin{figure}[b]
\includegraphics[angle=0,width=0.5\textwidth]{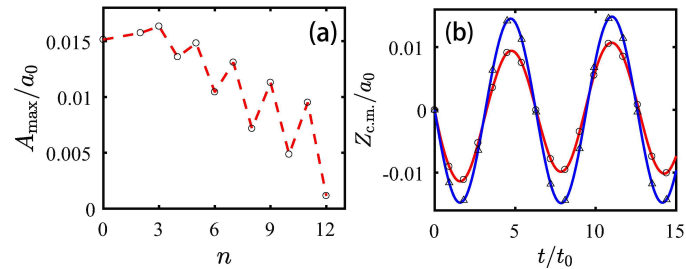}
\caption{(Color online) (a) The variation of the maximal amplitude $A_{\rm {max}}$ of the center-of-mass oscillation with respect to the wave number $n$ of the Kelvin wave perturbations. (b) The trajectories of the center-of-mass oscillation in the $z$-direction of the condensate containing a vortex ring with topological charge $s_j=2$ (red solid line with circles) and $s_j=1$ (blue solid line with triangles). All the other parameters used are the same as those in Fig.\,\ref{fig7}.}
\label{fig9}
\end{figure}

In Fig.\,\ref{fig9}(a), we plot the maximal amplitude of the dipole oscillation with respect to the wave number $n$. The zigzag line shows that the variation of the amplitude depends on the parity of the wave number of the Kelvin wave perturbations, which coincides with our analysis for Fig.\,\ref{fig8}. When the amplitude of Kelvin wave perturbation is smaller than 4, the maximal amplitude $A_{\rm {max}}$ for odd $n$ is always smaller than the subsequent even $n$. As shown in Fig.\,\ref{fig9}(b), we find that the vortex ring with higher topological charge ($s_j=2$) can also induce dipole oscillation of the condensate, but the amplitude of the dipole oscillation is depressed comparing with that of the $s_j=1$ case.

\section{CONCLUSIONS}\label{conclusion}

In this paper, we numerically study the collective dipole oscillation of a 3D BEC induced by an imprinted vortex ring and its deviation induced by Kelvin wave perturbations on the ring. As predicted, the collective excitation mode when a vortex ring is created in a condensate is nothing but the dipole oscillation of the center-of-mass of the condensate along the moving direction of the vortex ring ($z$-direction in our case). This excitation mode will not change with respect to the geometry of the condensate (trap aspect ratio), interatomic interactions and the initial radius of the imprinted vortex ring. However, the amplitude of the oscillation is very sensitive to these factors. There are two special radii of the vortex ring {to identify three dynamical modes of the oscillation of the center-of-mass of condensate and the vortex ring}. One makes the oscillation of the ring with respect to the centre-of-mass of the condensate disappear {(integrated mode with $R_v=R_{eq}$)}, and the other one makes the oscillation of the center-of-mass of the condensate cloud stops, respectively. {For all other values of initial $R_v$, the oscillation of the center-of-mass of the condensate is with smaller amplitude and period than the oscillation of the vortex ring.} We find there is a linear relation between the TF radius and the initial radius of vortex ring for making the condensate to be at rest (no center-of-mass oscillation). When the initial radius of the vortex ring is away from the equilibrium radius, the ratio between the amplitude of the center-of-mass oscillation of the condensate and that of the vortex ring itself decays rapidly close to zero.

When the Kelvin waves are introduced, the dynamics of the condensate and the vortex ring is quite different. If the amplitude of the added Kelvin waves is relative small, the dipole oscillation of the center-of-mass of the condensate will last but with a smaller amplitude, which shows the robust of the dipole excitation mode. We find that the amplitude of the dipole oscillation is sensitive to the modes of the Kelvin perturbation (planar, vertical and helical), and the parity of the wave number of the Kelvin waves. When the amplitude of the Kelvin perturbations is relatively small comparing with the initial radius of the vortex ring, the contribution of the depression of the dipole oscillation is mainly from the vertical Kelvin waves for even wave numbers, wile it is mainly from the planar Kelvin waves for odd wave numbers. The depression is greater for an even wave number than for the subsequent odd number, which gives a zigzag curve in the $A_{\rm {max}}$-$n$ plane. Increasing the topological charge of the vertex ring can also depress the dipole oscillation of the condensate induced by the vortex ring. When the amplitude of the the Kelvin perturbations is strong, the dipole mode can be destroyed due to the broken of the vortex ring structure.

\section*{Acknowledgments}

This work is supported by the National Natural Science Foundation of China under grants Nos. 11775178, 11775177, 11425522, 11434015 and 61835013, National Key R\&D Program of China under grants No. 2016YFA0301500, the Strategic Priority Research Program of the Chinese Academy of Sciences under grants Nos. XDB01020300 and XDB21030300, the Major Basic Research Program of Natural Science of Shaanxi Province under grants Nos. 2017KCT-12 and 2017ZDJC-32. This research is also supported by The Double First-class University Construction Project of Northwest University.




\end{document}